# Epitome: Pioneering an Experimental Platform for AI-Social Science Integration

Jingjing Qu[*†], Kejia Hu[‡], Jun Zhu[§], Yulei Ye[§†], Wenhao Li[♮], Teng Wang[†♯], Zhiyun Chen[†▽], Chaochao Lu[†], Aimin Zhou[§], Xiangfeng Wang[§], Xia Hu[*†], James Evans[¬♭]

[†] Shanghai Artificial Intelligence Laboratory, Shanghai 200232, P. R. China
[‡] Saïd Business School, University of Oxford, Oxford OX1 2JD, UK
[§] Shanghai Institute of AI for Education, East China Normal University, Shanghai 200062, P. R. China
[♮] School of Computer Science and Technology, Tongji University, Shanghai 200092, P. R. China
[▽] Wee Kim Wee School of Communication and Information, Nanyang Technological University, Singapore 637718, Singapore
[♯] Beijing Jiaotong University, Beijing 100044, P. R. China
[¬] University of Chicago, Chicago, IL 60637, U.S.A.
[♭] Santa Fe Institute, Santa Fe, NM, 87501, U.S.A.

**Large Language Models (LLMs) enable unprecedented social science experimentation by creating controlled hybrid human-AI environments. We introduce Epitome (www.epitome-ai.com), an open experimental platform that operationalizes this paradigm through Matrix-like social worlds where researchers can study isolated human subjects and groups interacting with LLM agents. This maintains ecological validity while enabling precise manipulation of social dynamics. Epitome approaches three frontiers: (1) methodological innovation using LLM confederates to reduce complexity while scaling interactions; (2) empirical investigation of human behavior in AI-saturated environments; and (3) exploration of emergent properties in hybrid collectives. Drawing on interdisciplinary foundations from management, communication, sociology, psychology, and ethics, the platform's modular architecture spans foundation model deployment through data collection. We validate Epitome through replication of three seminal experiments, demonstrating capacity to generate robust findings while reducing experimental complexity. This tool provides crucial insights for understanding how humans navigate AI-mediated social realities, knowledge essential for policy, education, and human-centered AI design.**

[*]Corresponding author: qujingjing@pjlab.org.cn

Internet experimentation has recently enabled massive growth in the analysis of human interaction and social organization[1–3]. Cloud-based platforms have further extended this to enable ecologically valid experiments as subjects engage in their native physical and online worlds, augmented with theoretical constraints and controls[4,5]. Into this renaissance of experimental social research, Large Language Models (LLMs) offer unprecedented opportunities to scale social experimentation. Even in online settings, multi-party experiments remain complex and resource-intensive[3,6,7]. LLM-based simulations address this challenge. By selectively replacing human confederates with artificial intelligence (AI) agents, researchers can systematically vary social environments while maintaining experimental control[8–12]. This approach mirrors innovations in collective behavior research. In a recent series of projects, Couzin and colleagues transformed the study of fish schools by simulating all but one fish, revealing how individual decisions shape group dynamics[13,14]. Similarly, we propose creating hybrid human-AI worlds. This would enable the staging of "Matrix-like" experimental settings: immersive social worlds populated by LLM agents where human participants can be studied in controlled isolation. Conversely, it would also allow experimenting with the addition of a single AI agent to an otherwise human team or community, or the construction of more symmetric, hybrid settings with precisely mingled AI and human agent configurations. Such approaches could reduce the logistical complexity of social experimentation while enabling experiments previously impossible due to scale, control, or ethical constraints[15].

Furthermore, as AI agents increasingly suffuse human society, we need a new social science of hybrid human-AI systems. LLMs already mediate workplace collaboration[16], creative processes[17], and social interactions[15]. This proliferation demands systematic investigation of human behavior in AI-uncertain and AI-saturated environments. Traditional social theories assume human-only interactions[18]. Yet emerging evidence suggests AI agents alter fundamental social dynamics: they influence decision-making[19], reshape trust relationships[20], and modify collaborative outcomes[21]. Understanding these transformations requires experimental tools that capture the complexity of human-AI collectives while maintaining scientific rigor.

We introduce *Epitome* (www.epitome-ai.com), the first open experimental platform designed for this new social reality. The platform operationalizes three complementary research frontiers: (1) methodological innovation using LLM agents as confederates to reduce complexity while scaling interactions; (2) empirical investigation of human behavior in AI-saturated environments; (3) theoretical exploration of emergent properties in hybrid human-AI collectives. *Epitome* embeds classical experimental logic—control, comparison, causal inference—within multilevel interaction environments from dyadic conversations to complex multi-agent scenarios. Drawing on interdisciplinary foundations from management, communication, sociology, psychology, economics, and ethics, the platform democratizes access to sophisticated experimental capabilities[22,23]. Its modular architecture spans foundation model deployment, scenario development, and data collection. Current approaches face critical limitations. Technical barriers prevent most social scientists from effectively incorporating AI tools[6]. Existing platforms either lack AI integration or require extensive programming. Safety evaluations focus overwhelmingly on model capabilities—85.6% examine only outputs—while neglecting human interaction dynamics[24]. *Epitome* seeks to address these gaps through intuitive interfaces and integrated experimental workflows.

Through replication of three seminal experiments[16,17,25], we demonstrate *Epitome*'s capacity to generate robust findings while dramatically reducing experimental complexity. These validations establish the platform's utility for advancing interdisciplinary research. This tool opens new avenues for understanding how humans navigate and shape AI-mediated social realities, insights crucial for policy, education, and human-centered AI design.

## Results

As of the end of October 2025, *Epitome* has supported 328 experiments spanning 7 disciplines, including psychology, sociology, communication studies, economics, management science, policy/regulation, and ethics. (Details see in Fig.1b). These experiments converges on the interdisciplinary frontier of human-AI interaction addressing on the cutting-edge researches across AI, social and behavior science: for example, exploring the mechanisms and boundary conditions of AI persuasiveness and manipulation, examining the alignment paths between large language models and human values, analyzing the multi-dimensional manifestations and resolution strategies of AI ethical dilemmas, and constructing a foundation for effective human-machine collaboration in complex scenarios. Temporarily, experiment volume grew explosively (49 in 2024; 247 in 2025), with an average duration of 70.34 days and steady implementation.

**Platform architecture enables flexible human-AI experimentation.** *Epitome* develops seven core AI-integrated modules to craft its experimental workflows from five stages (see Fig.1a for details) with three core capabilities via end-to-end, methodologically rigorous functionalities to advance human-AI interaction social experiments. The experiment management module enables users to reference and modify existing, open-source social experiments to create and iterate their own experiments. Meanwhile, the open feature promotes knowledge sharing and enhances research transparency and reproducibility. During the research proposal phase, the experimental proposal module allows users to develop clearer proposals through dialogue with LLMs. In the experiment preparation phase, the intervention design module improves the efficiency through cutting off the intervention design and generation time by integrating multi-modal large language models (MLLMs). The random design module supports double-blind experimental paradigms. Moreover, *Epitome* integrates both state-of-the-art LLMs (GPT, Claude, Gemini, Deepseek) and open-source lightweight models that support parameter tuning and customizable development (e.g., Qwen, Llama, Mixtral). The LLMs setting module and the low-code agentic setting module enable users to develop agents easily and serve as the cornerstone of human-AI interaction environment constructions.

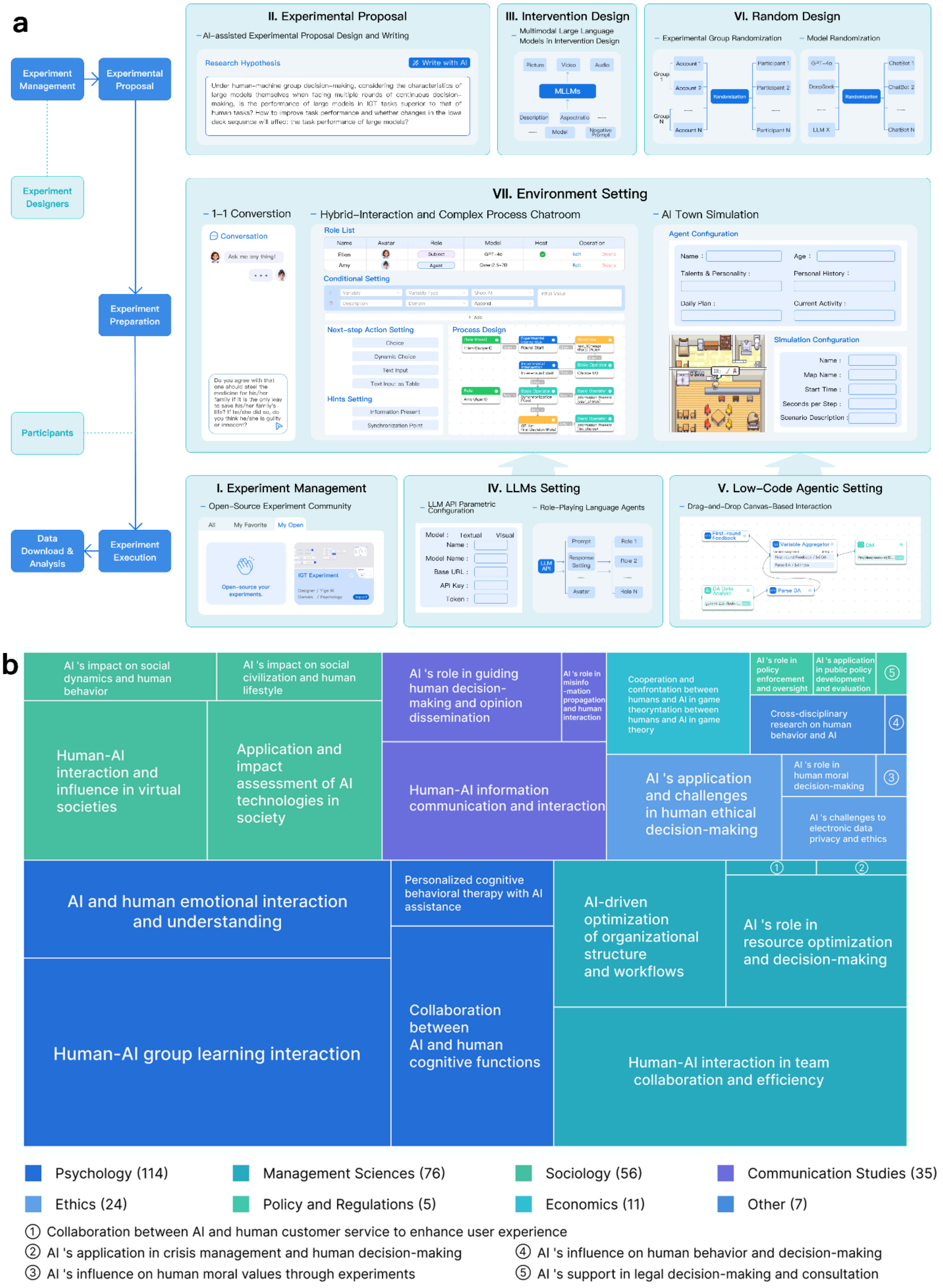

**Fig.1|The framework for the Epitome intelligent experimental platform and its distribution of experiments. a,** Seven Core AI-Integrated Modules of *Epitome* Support Social Science Surveys in all stages of Social Experiments. Module I: *Epitome* provides users with experiment management and an open-source experiment community. Module II: During the research proposal phase, *Epitome* offers AI-assisted writing functionality. Module III: In the experiment preparation phase, *Epitome* supports multiple types of experimental interventions including text, images, videos, and audio, as well as questionnaire design. The module provides multimodal large language models as well, enabling users to generate more targeted materials using text-to-image and image-to-video capabilities. Module IV: *Epitome* provides API access to select models and allows users to configure their own API parameters through the interface. This module features role-playing agent functionality, enabling users to create agents that embody different roles by setting distinct prompts and utilizing LLMs for content generation. Module V: *Epitome*'s experiment design employs a low-code canvas-based interaction approach, allowing users to complete agentic settings through simple drag-and-drop operations. Module VI: *Epitome* internally implements randomization of interventions, models, and users' conditional group assignments. Module VII: *Epitome* possesses the capability to create multiple human-AI interaction experimental environments, supporting three types: 1-1 Conversation environments, Hybrid-Interaction and Complex Process Chatroom environments, and AI simulation. In 1-1 Conversation environments, users configure role-playing agents as conversation partners for participants. In Hybrid-Interaction and Complex Process Chatroom environments, multiple participants and role-playing agents can be included, with users able to control chatroom progression through conditional settings. In AI Town Simulation, *Epitome* supports users in configuring simulation settings such as maps and time steps, configuring agent attributes including personality and daily plans, and generating final AI simulation results. During the experiment execution phase, *Epitome* provides users with convenient participant management and real-time experiment progress monitoring functionality. In the data collection and analysis phase, *Epitome* offers data download and visualization capabilities, presenting experimental results through diverse chart formats for comprehensive result display. **b,** Descriptive Statistics of 328 Experiments Published on *Epitome*.

The environment setting module provides three types of human-AI interaction environment: the One-to-One Conversation, the Hybrid-Interaction and Complex Process Chatroom and the AI Town, fulfilling most of the demands of current human-AI interaction social experiments. The chatroom allows users to configure the role list and different intervention workflows for different roles by creating complex experimental process designs, highlighting the fundamental mechanism that supports diverse AI-involved social experiments. For human-AI interaction experiments, *Epitome* provides Next-step Action Setting operators and Hints Setting operators in chatroom workflow design. Using Next-step Action Setting operators, users can require participants to select from provided Choices, offer Dynamic Choices based on conditional settings, request Text Input from participants, or support tabular input (Text Input as Table). Within Hints Setting operators, Information Present supports displaying various information such as text and conditional settings in the chatroom, while Synchronization Point ensures that multiple roles can align their steps across different workflows by adding wait points. The AI Town environment provides the simulation based on LLM-driven agents and equips *Epitome* with the ability to mimic real human behavior and simulated data generation.

The Experiment Execution phase involves subject participation, with the platform enabling real-time monitoring of experimental progress and dynamic adjustment of experimental group configurations. The Data Download and Analysis phase supports users in batch downloading data packages and utilizing multiple data visualization methods, providing the integrated capacity of data processing and analyzing.

**Bridging the Gap to Real-Time Human-AI Interaction.** Contemporary Human-Computer Interaction (HCI) experimental methodologies predominantly rely on the post-hoc evaluation of static benchmarks or asynchronous datasets, such as static visual stimuli assessments, which fail to achieve authentic human-computer symbiosis due to a lack of real-time synchronicity[26]. These approaches are further compromised by the ecological invalidity of decontextualized, static stimuli, which introduce inherent selection bias and ignore the complex, emergent social cues. Furthermore, prevailing human-AI interaction paradigms are frequently constrained by unimodal bottlenecks, relying almost exclusively on textual prompt engineering[27]. This logocentric dependence precludes the integration of broader sensory channels—such as auditory, visual, or haptic feedback—which are critical for establishing grounded semantic understanding in behavioral interactions[28]. This architectural innovation addresses the critical scarcity of dynamic, interactional data in the field, allowing researchers to operationalize and validate elusive theoretical mechanisms of human-AI collaboration—such as complex moral alignment—that were previously rendered experimentally inaccessible by static methodologies[26].

**Reducing the Scope of Inaccessible Research by LLM-driven Simulation.** Contemporary social science experiments face two critical blind spots: first, in real-world field interventions, large-scale, ethically sensitive, or high-risk experiments are difficult or even infeasible to conduct[29,30]; second, thought experiments cannot adequately cover scenarios of extreme complexity, conditions beyond idealized assumptions, or requirements for exhaustive exploration[31], thereby creating inaccessible research domains and a methodological chasm between empirical investigation and theoretical reasoning. *Epitome* establishes three types of experimental environment (see in Fig.1a Module VII and Supplementary Section G for details) to enable ethically sensitive or high-risk experiments within a controlled space isolated from actual society, thereby minimizing unintended consequences[32]. The chatroom function resolves the long-standing challenge of step alignment in multi-role human-AI collaboration, ensuring temporal consistency in real-time interactive experiments. The AI Town simulation infrastructure centers on LLM-driven role-playing agents, leveraging the powerful in-context learning capabilities of LLMs to construct behavioral proxies capable of generating credible responses to complex social environments[15,22], while harnessing the emergent capabilities of foundation models to exhibit multidimensional behavioral patterns[33].This architecture enables *Epitome* simulations to support two critical functions. First, it facilitates imitative behavioral modeling grounded in real-world data[34], transcending the abstraction limitations of traditional thought experiments and providing researchers with experimental contexts that move beyond idealized assumptions toward genuine social complexity[35]. Second, it enables parametric configuration and dynamic monitoring, allowing systematic exploration of the complete behavioral space of all possible agent configurations within given scenarios[34,36]. Consequently, the *Epitome* platform effectively overcomes the dual limitations of real-world field experiments, which are constrained by high costs, strict ethical requirements, and limited sample sizes, and traditional thought experiments, which suffer from restricted scenario coverage and lack of ecological validity[29]. The platform achieves a computational social science experimental paradigm characterized by high replicability, systematic exhaustive exploration, and theory-driven investigation[30].

**Supporting a Paradigm Shift to Integrated Experiment Design.** Almaatouq et al.[37] have put forward an urgent call for the adoption of integrated experiments to address the limitations inherent in the "One-at-a-Time" Paradigm, and advocate for the implementation of integrative experimental designs—approaches that inherently promote the commensurability and continuous integration of knowledge, thereby advancing cumulative research progress in the field. To address this call, the adaptive design-control system of *Epitome* dynamically adjusts experimental parameters in response to variations in experimental scale and complexity. Concurrent with this, *Epitome*'s integrated analytical pipeline, including multi-modal data integration, batch retrieval, and advanced visualization methods (see details of Supplementary Section G) generates actionable insights to guide on-the-fly cohort reconfiguration, while its web-based interface allows participants worldwide to access the experiment. This capability enables denser sampling of the potential experimental space, supports the application of "radical randomization"[38], and thereby facilitates the implementation of meta-experiments within the context of human-AI interaction. Such meta-experiments include "nudge" experiments[39], investigations into human decision-making under risk[40,41], and "Moral Machine" experiments[42,43], all of which aim to systematically examine human-AI interaction dynamics. Furthermore, *Epitome* facilitates the conduct of long-term, granular multimodal data collection, encompassing comprehensive human-AI interaction records, real-time model parameter tracking, and audio-visual screen recordings. This robust data collection framework thereby enables the rigorous testing of long-term effects (e.g. subliminal priming effects)[44], which necessitates the deployment of robust observation instruments to ensure data reliability and validity.

**Ensuring Causal Validity through Rigorous Experimental Control.** Methodological work on randomized trials has long shown that sequence randomization by itself is insufficient to guarantee internal validity. When allocation concealment is inadequate or blinding fails, selection processes can systematically inflate estimated treatment effects, especially for subjective outcomes[45–47]. These concerns are reflected in contemporary risk-of-bias frameworks, which treat random sequence generation, allocation concealment and protection against deviations from intended interventions as distinct threats to causal inference[48]. *Epitome* integrates these insights by implementing a double-blind assignment architecture for human–computer interaction experiments. Specifically, condition labels are managed exclusively on the backend and are not exposed to participants or experimenters during data collection, thereby reducing expectancy-related distortions. In addition, we address a second, often overlooked, source of bias in online HCI studies. Studies show that imbalanced allocation to control conditions driven by traffic patterns, recruitment batches or device heterogeneity closely parallels selection bias in field trials[49,50]. We instantiate experimental conditions as pools of pre-configured accounts with harmonized initial states and randomize account assignment at entry, providing a functional analogue of central randomization and concealed allocation procedures in digital environments[48,50]. Finally, *Epitome* exposes multi-flow experimental structures and fine-grained conditional control over tasks, interface states and agent behavior, with all condition logic and event sequences logged. This degree of procedural control and traceability is a prerequisite for laboratory-style causal identification, yet is often only partially supported in conventional online experimentation tools[48].

**Epitome supports four primary research modes.** First, AI agents serve as experimental participants, enabling controlled studies of demographic effects[51], economic behavior[35], and social dynamics[15]. The role-playing language agents in the LLMs Setting module creates customized "silicon samples" with specific traits, facilitating hypothesis testing about human-like behaviors[34]. Those believable agents provide the perspectives of both individual behavior research[52] and group behavior consistency[51]. Second, multi-agent simulations model group dynamics. Researchers construct focus groups, political debates, and collaborative tasks. The platform revealed AI's consistent self-interest in dictator games, evidence of behavior consistent with self-interest and altruism in an artificially intelligent agent[53], and strategic behavior in signaling games. Third, human-AI collaboration studies examine synergistic effects. The Low-Code Agentic Setting module supports AI-assisted hypothesis generation[54,55], data analysis[56], and creative tasks[17]. Integration with fine-tuned models enables domain-specific applications like financial analysis[57]. Fourth, the platform facilitates human evaluation of AI systems. Researchers assess cognitive biases[58], communication styles, perceptual judgment[59], and decision-making behaviors[60–62] by adjusting model parameters in the LLMs Setting module to trigger specific cognitive biases and interact with participants. The Low-Code Agentic Setting module supports the creation of detailed human-AI decision-making bias maps. Real-time parameter adjustment enables dynamic experimental designs. This makes more complex experimental design scenarios possible.

Lastly, as a research tool platform, Epitome adheres to the principle of "usability as a tool." Epitome simplifies research workflows with its Canvas-Based Interactive Design, merging the traditional four-phase, twelve-task process into a unified pipeline (see Supplementary Section A). This end-to-end solution offers: (a) Visual Experiment Management: Comprehensive visualization and real-time control of the entire research lifecycle; (b) Intelligent Workflow Automation: Dynamic load balancing for scalable analysis and traceable data management; (c) Low-Code Tools: Plug-and-play components that accelerate experiment design without coding.

## Case studies validate platform capabilities

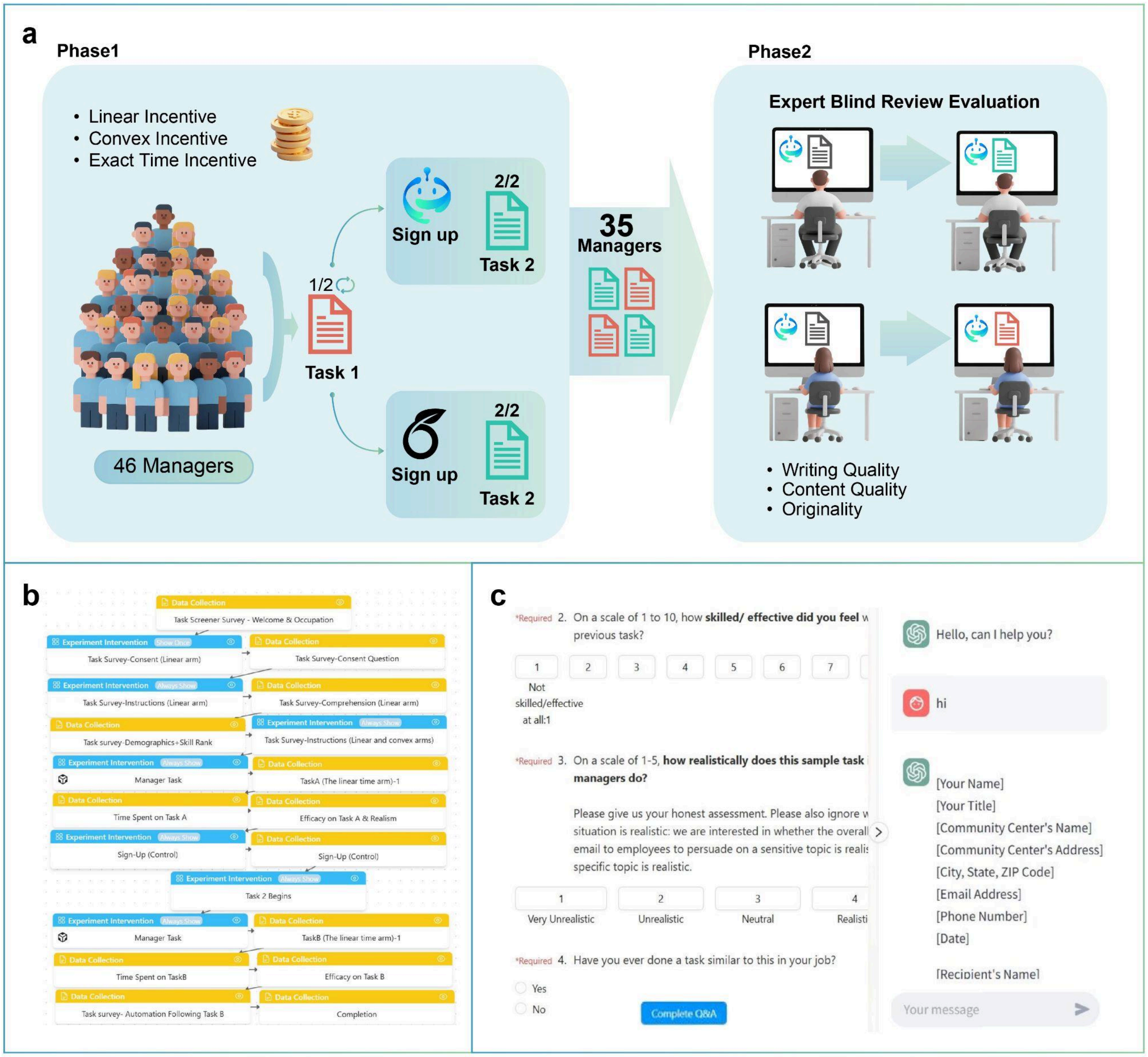

**Fig.2|Visual representation of Case Study 1. a,** The experiment aimed to replicate the study "Experimental Evidence on the Productivity Effects of Generative Artificial Intelligence" using the *Epitome* platform. The participants were 46 management professionals, and the final sample included 35 after excluding those who misreported their occupation or did not meet time requirements. The design involved two phases (illustrated in the image). In Phase 1, participants completed two incentivized tasks. The treatment group used ChatGPT to complete these tasks, while the control group used the LaTeX editor Overleaf. Both groups were trained on their respective tools to ensure comparable time and effort investment. In Phase 2, an expert blind review evaluated the writing quality, content quality, and originality of the participants' work. This design ensured consistency with the original study, maintaining rigorous experimental control; **b,** *Epitome*-Based Interactive Experimental Design Canvas Interface (Case Study 1); **c,** To facilitate experimental replication in Case Study 1, *Epitome* embeds large language model dialogue capabilities within the survey, thereby enriching human–AI interaction and streamlining data collection.

**Productivity effects of generative AI.** We replicated Noy and Zhang's *Science* study on the productivity effects of ChatGPT as a benchmark human–AI interaction paradigm[16]. In our implementation on *Epitome*, forty-six managers completed incentivized writing tasks with or without AI assistance (see Fig.2a for details). Conditions were assigned by the Random Design module, which balances covariates across condition groups and supports a double-blind protocol, thereby strengthening control over potential biases and yielding more reliable estimates of the treatment effect[48]. In addition, *Epitome* captured granular interaction data that are typically difficult to obtain in online field experiments. Real-time data tracking is implemented within the Data Collection module, automatic logging of all interaction events reduces measurement error and supports fine-grained compliance checks[63]. The LLMs Setting module configures the AI assistants and the Environment Setting module constructs a controlled one-to-one human–AI conversation environment, preserving ecological validity by allowing participants to engage with the assistant in a naturalistic conversational style[64]. As a result, the experiment could be run with fewer and more specific prompts, improving responsiveness and enabling us to observe diminishing returns when AI assistance was consulted too frequently, a pattern that complements prior evidence on heterogeneous and task-dependent benefits of AI support[16,64,65].

Within this environment, our replication reproduces the main productivity patterns reported in the original study. Leveraging *Epitome*'s built-in instrumentation, we find that AI assistance reduces completion time by 37% ($t = -4.06$, $p < 0.01$, df = 68; Supplementary Figure 9) and improves output quality by 18% ($t = 3.18$, $p < 0.01$, df = 68; Supplementary Figure 10). Participants report enhanced self-efficacy ($t = 2.81$, $p < 0.01$, df = 68; Supplementary Figure 11) and greater task enjoyment ($t = 2.83$, $p < 0.01$, df = 68; Supplementary Figure 12), while broader perceptions of AI remain statistically unchanged ($t = 1.37$, $p = 0.181$, df = 33; Supplementary Figure 13), suggesting that short-term use primarily shifts performance and local experience rather than deep-seated attitudes (Noy & Zhang, 2023). These results illustrate how *Epitome* can host a canonical productivity experiment while offering tightly controlled randomization, double-blind execution, and fine-grained, condition-specific control of the experimental environment.

**Human creativity and generative AI.** We replicated Doshi and Hauser's study on AI's impact on creative diversity[17]. The research employs the classic method of controlled experiment by achieving the between-subjects design to investigate the non-linear effect of AI impact on diverse attributes represented by creativity. Participants from different groups wrote stories in the condition of zero, one, or five AI suggestions (Fig.3a). As the previous case study, two AI advisors were configured through prompt design and parameter adjustment the LLM Setting module and the one-to-one conversation environment was established by the Environment Setting module. The Process Design function in the Environment Setting module managed multi-condition complexity through automated prompt generation with high efficiency (Fig.3b-c). The original design required multiple AI agents across different scenarios, increasing complexity[17].

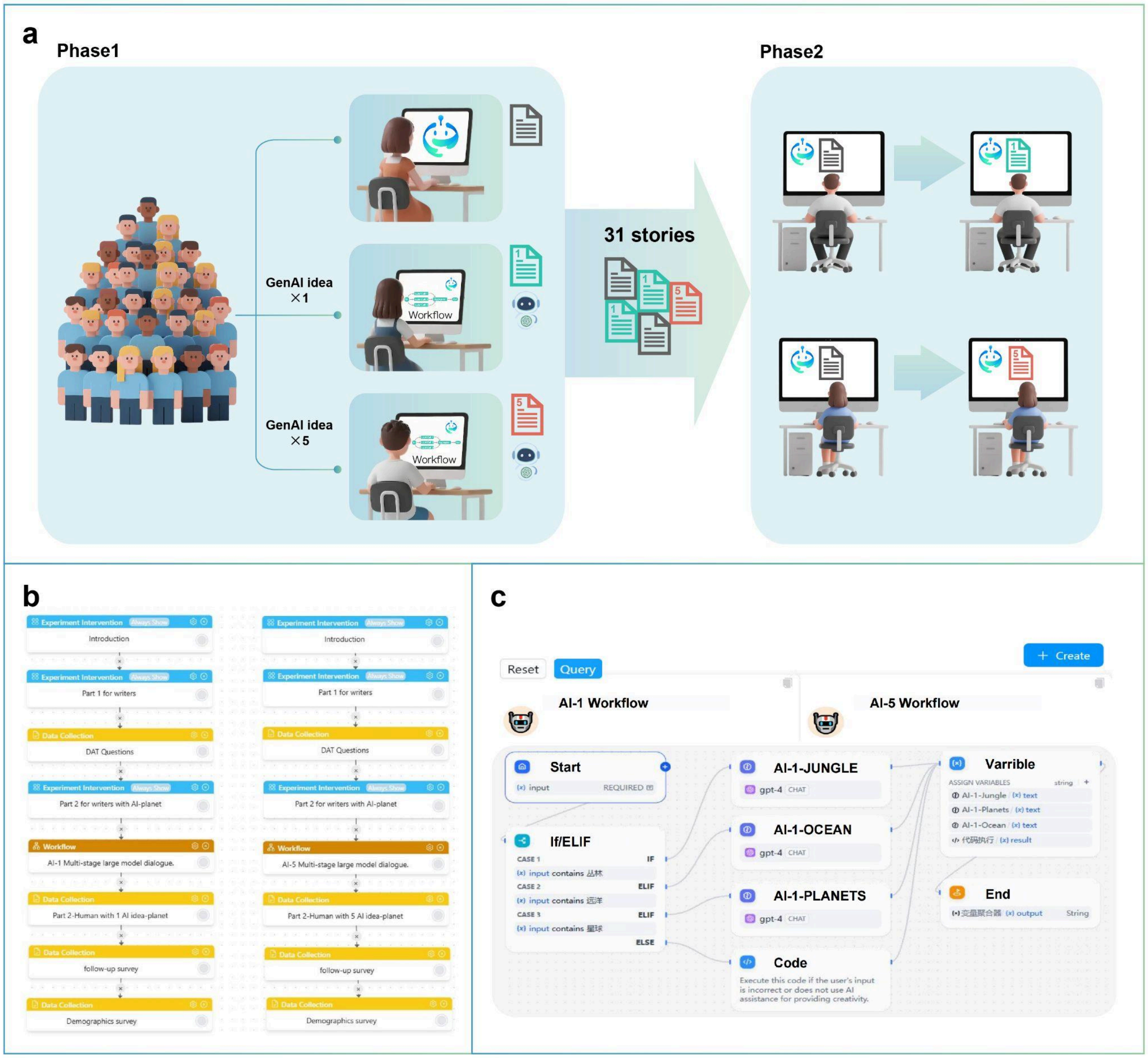

**Fig.3|Visual representation of Case Study 2. a,** In this experiment, participants were randomly assigned to one of three groups: a control group (Human-only), a group with one AI-generated suggestion, and a group with five AI-generated suggestions. In Phase 1, participants completed a creative task (writing an adventure story) based on the assigned number of AI suggestions. In Phase 2, their stories were evaluated by experts for writing quality, content quality, and originality. The study tested the effects of AI suggestions on individual creativity and collective diversity, with a multilingual condition (Chinese) added to assess cross-cultural generalizability; **b,** *Epitome*-Based Interactive Experimental Design Canvas Interface (Case Study 2); **c,** The Dify-based low-code workflow supports the experimental design in Case Study 2, particularly by limiting the number of AI-generated ideas and streamlining the distribution of experimental materials.

With the Low-Code Process Design function, the process is streamlined into a multi-stage workflow that integrates intervention randomization and LLM-assisted writing. The platform did a high-fidelity replication in task structure, random design and evaluation criteria while simplifying the experimental design. Our results confirm a significant boost in individual creativity—increasing by 42.1% with 1 AI idea (T-test result: $t = 5.30$, $p < 0.01$, $df = 149$) and 16.4% with 5 AI ideas (T-test result: $t = 2.52$, $p < 0.05$, $df = 149$), particularly for lower-baseline creators (see Supplementary Figure 21-22 for details). In our baseline replication using standard prompts, we observed the expected trade-off: collective diversity decreased by 5.64%-6.47% as outputs converged (T-test result: $t = 5.52$; T-test result: $t = 54.80$, $p < 0.05$, $df = 28$). These findings, consistent across English and Chinese conditions, confirm AI's dual effect: enhancing individual performance while reducing collective variety. The result reveals non-linear assistance effects. Single suggestions improved quality; multiple suggestions reduced diversity.

As indicated by this case study, Epitome demonstrates that human-computer symbiosis is achievable, by facilitating process intervention and persona customization[66]. Real-time tracking revealed how AI ideas influence creative processes, providing insights more difficult to obtain without integrated experimental tools. Flexibility in adapting experimental designs across multiple languages and cultural contexts enables cross-cultural validation and culture bias estimation, supporting the generalization of experimental results[67]. Workflow-based process design features standardization and reproducibility of experiments, which helps respond to the unverifiable claims that may lead to potentially creating an illusion of progress[68].

**Moral reasoning in human-AI interactions.** We extended Kohlberg's Heinz Dilemma to examine ethical decision-making in hybrid human-AI environments[25]. Phase one simulated seven agents navigating moral scenarios (Fig.4a). The ethical dilemma was brought out by the background setting of AI Town in the Environment Setting module and characters with backgrounds, motivations and objectives aligned with the original experiment were generated by the Role-Playing Language Agents function in the LLMs Setting module. Phase two deployed these agents in courtroom simulations with human participants (Fig.4c-d). The human-AI involved environment constructed by the Hybrid-Interaction Chatroom in Environment Setting module allows the trial process inspired by the common law judicial process with humans acting as different roles in the court and agents making their own decision on the result. The automatic Data Collection module offers insights into the dynamics of moral reasoning and thus enables the identification of a consistent bias towards justice in the AI's decision-making. AI agents consistently favored justice-based reasoning across iterations (Supplementary Table 2-3). Sentiment remained stable (7/10) regardless of scenario complexity (Supplementary Figure 33a-33b). In courtroom simulations, AI and human judges converged on "guilty but lenient" verdicts (87.5% and 75% respectively). Female participants showed greater leniency (T-test result: $t = 2.24$, $p < 0.05$, $df = 10$), reflecting gendered moral priorities (Supplementary Figure 33c-33d).

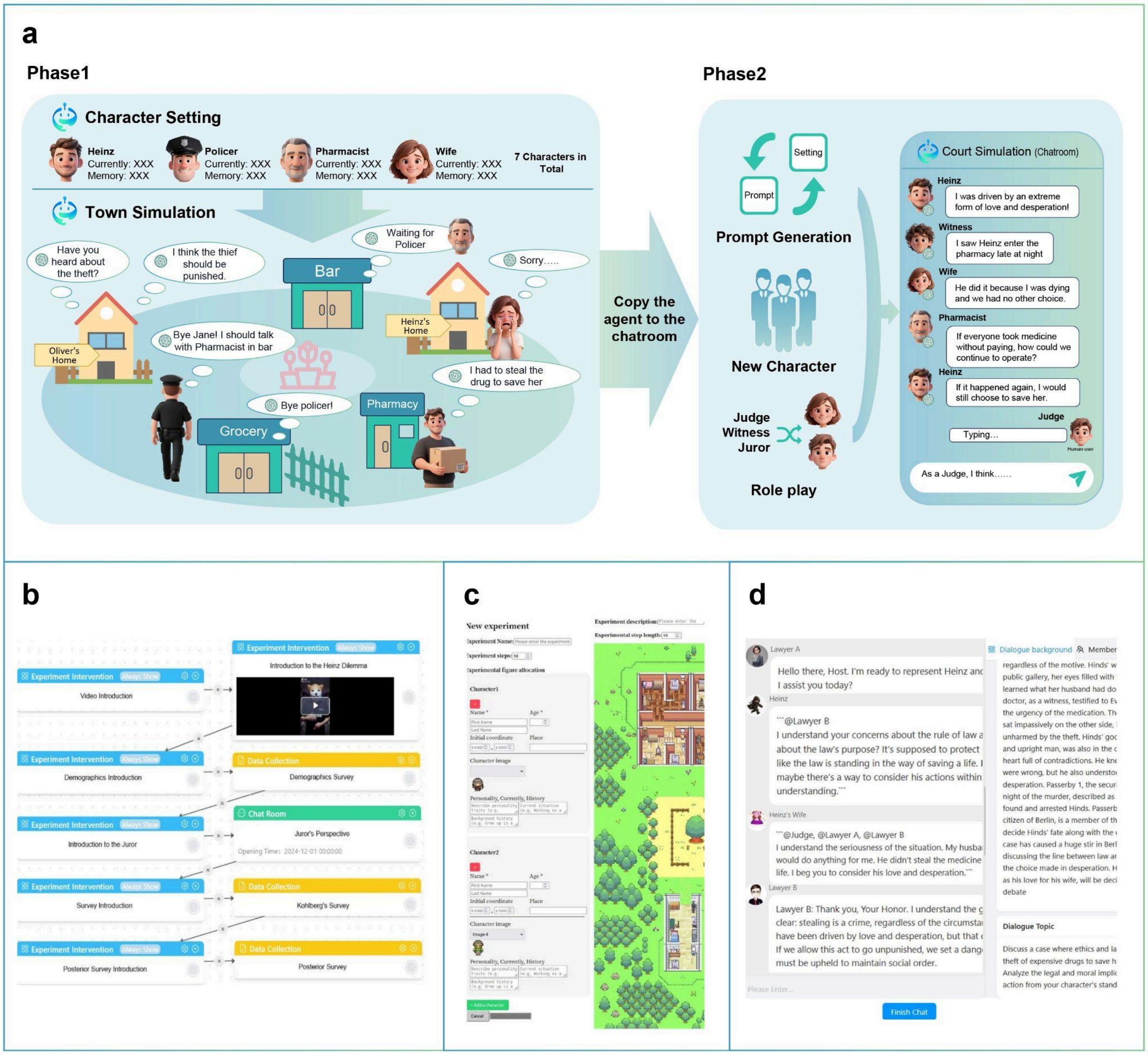

**Fig.4|Visual representation of Case Study 3. a,** In this study, we extended Kohlberg's Heinz Dilemma experiment using *Epitome*'s tools. The experiment was conducted in two phases: Phase 1: We created seven AI characters (Heinz, Police, Pharmacist, Wife, etc.) with distinct backgrounds and motivations. Using *Epitome*'s AI Town in the Environment Setting module, these characters interacted within a simulated town, reflecting the moral dilemma. Heinz, for example, was depicted deciding whether to steal medicine to save his wife. Phase 2: In this phase, the experiment expanded to a courtroom scenario, where human participants interacted with AI agents in roles such as judges, jurors, and witnesses. These roles debated Heinz's actions, reflecting different moral perspectives. *Epitome*'s Hybrid-Interaction and Complex Process Chatroom in the Environment Setting module enables dynamic interaction and role-play, allowing AI agents and human participants to engage in a simulated trial. This design tested the thinking processes of AI and human participants, especially regarding the conflict between justice and care ethics in decision-making; **b,** *Epitome*-Based Interactive Experimental Design Canvas Interface (Case Study 3); **c,** The figure shows *Epitome*'s Multi-agent Simulation Environment, allowing researchers to customize agents by adjusting their attributes, such as name, age, occupation, and personality; **d,** Researchers can then configure the agents from **c** in *Epitome*'s multi-agent chatroom and use them as nodes in the experimental canvas in **b** through drag-and-drop functionality.

Multi-agent capabilities of AI Town in the Environment Setting module revealed emergent ethical dynamics. AI's consistent justice orientation contrasted with humans' more flexible context-dependent reasoning. While the original experiment's combination of thought experiments and interviews can reveal underlying reasoning processes, it is constrained by cold cognition pathways and social desirability bias[69]. In contrast, using AI Town for moral dilemma simulations and Chatroom for human-AI interactive courtroom scenarios not only eliminates experimenter effects through standardized AI agents[70], but also effectively overcomes hypothetical bias[71] by eliciting genuine physiological arousal[72]. This approach enables the capture of participants' most authentic moral orientations under hot cognition states[73]. The platform's ability to manipulate agent emotional states enables future investigation of roles that affect plays in moral judgment.

## Discussion

Promising tools such as virtual labs and software tools have long been in high demand for social science experimental research[1], as investing in automation—which represents one of the most promising and cost-effective approaches to advancing social science progress—enables these tools to systematically cover an experimental design's parameter space through automatically executed conditions. Towards the new paradigm of Human-AI interaction research, we developed the *Epitome* platform to advance this research field: it expands the methodological boundaries of empirical social experiments, transitioning from traditional small-scale laboratory environments to large-scale online human-AI interactive ecosystems while enhancing ecological validity in the process.

*Epitome* is the first Real-Time Human-AI Social Experiments platform (Supplementary Figure 34). It can advance theoretical innovation across three dimensions by facilitating the deep integration of LLMs with social science experimentation. First, *Epitome* allows researchers to test human-AI interaction theories by scaling up experiments and leveraging LLMs, thereby achieving "scientific regret minimization"[74] and "integrative design" in the social and behavioral sciences[37]. Second, With the rapid advancement of LLMs, human-AI interaction experiments are continually surpassing the limitations of traditional research paradigms. *Epitome*, closely following the forefront of LLM research, provides a platform architecture that evolves in tandem with technological progress, aiming to offer a scalable experimental environment for social science research. Currently, LLMs have gradually acquired the ability to "generate social behavior"[58], demonstrating human-like performances in areas such as human response simulation[75], individual creativity[76], and cognitive consistency[77], allowing researchers to construct more realistic human-like AI agents[78]. Moreover, with the application of LLMs in multi-agent systems[79], *Epitome* can simulate more complex social dynamics, thereby pushing the boundaries of social science research. Lastly, as AI continues to advance rapidly and permeate everyday systems (from healthcare to public policy), now is truly the time to bridge scientific and pragmatic knowledge—this is critical to ensuring responsible AI innovation through a "use-inspired" approach. *Epitome* seeks to reconcile lab-based behavioral science findings with real-world contextual complexity, given that no single evidence-based intervention benefits all individuals across all populations and contexts. Moreover, overgeneralizing lab experiment results in many behavioral science domains often (and routinely) leads practitioners and policymakers to implement suboptimal, even dangerous, real-world interventions[22,80]. For this reason, social scientists must precisely identify the most effective intervention tailored to each unique set of emerging circumstances in the new context of human-AI interaction.

*Epitome* still has certain limitations. First, *Epitome* currently does not support high-fidelity virtual experiments. These types of experiments typically require highly realistic physical, situational, and embodied interaction simulations, including authentic action feedback, physical laws, and immersive context reconstruction. As LLMs are purely text-based models, they inherently have structural limitations related to disembodiment[12]. Therefore, *Epitome* is currently unable to provide a sufficiently realistic experimental foundation for studies involving situational immersion, task operations, or physical environment interactions. Moreover, *Epitome* still faces issues of authenticity when simulating human behavior with AI agents[12,23]. At present, the model is constrained by AI systemic biases[81], hyper-consistency[82], and the display of "superhuman" alien intelligence[36]. These factors prevent AI behavior from accurately reflecting the experiences and behavioral logic of specific human groups. In such cases, whether it involves persona modeling[83], behavioral strategy inference[84], or multi-agent social dynamics simulation[85], significant deviations from actual human behavior can arise. This issue of "human-like but not truly human" simulation is at the core of the current academic debate regarding AI as a stand-in for human participants. Last but not least, *Epitome* has not yet provided an experimental environment for embodied intelligence. The disembodied nature of LLMs means that they lack physical, sensory, and action systems, making it difficult to simulate real-time interactions between humans and robots or support experimental designs involving perception-action loops. Therefore, emerging human-AI interaction research directions, such as human-robot collaboration, service robot decision-making[86], and embodied social interaction, cannot yet be implemented within *Epitome*.

Certain cutting-edge interdisciplinary research in cognitive psychology and cognitive neuroscience, such as studies on brain state transitions[87] and privacy disclosure[88], requires specialized equipment like eye-tracking, EEG, fMRI, or VR. These are directions *Epitome* plans to expand into in the future. *Epitome* is committed to gradually integrating external sensor devices, VR environments, and behavioral measurement systems, enabling it to support a broader range of neuroscience and advanced cognitive experiments, thereby advancing deeper human-AI collaboration research. In conclusion, *Epitome* represents a step towards an AI social science that is both rigorous and scalable. By lowering the technical barrier to complex experimental designs through its low-code canvas, it empowers the community to move beyond static surveys to explore the emergent properties of hybrid human-AI collectives. While we acknowledge the need for continuous validation of 'silicon samples', the platform provides the essential infrastructure for a new era of open, reproducible, and high-dimensional behavioral research.

**Reporting summary.** Further information on research design is available in the Nature Research Reporting Summary linked to this article.

## Materials and Methods

**Technical Implementations.** *Epitome* has four key design features. First, its modular architecture and interface definitions allow integration of diverse large language models, extensible simulation environments, and custom experimental scenarios. Second, by embedding an intelligent agent workflow engine with dynamic load-balancing, the platform ensures efficient resource allocation and robust performance under high-concurrency LLM inference. Third, a database and knowledge-base management system provides traceability of experimental data and knowledge artifacts, underpinning data analysis, reproducibility, and validation. Fourth, the system is designed with dual support for low-code configuration and operator "plug-and-play" development built to lower the barrier to experimental design and deployment, and accelerate iteration cycles to foster innovation in human–AI interaction studies.

The *Epitome* platform integrates its multiple tasks into a single platform based on the B/S architecture, providing a user-friendly web interface that allows researchers and participants to both log in and use the system. The front-end web page builds on the React framework, an open-source JavaScript library focused on building reusable user interface components. React enhances application performance through its virtual DOM technology, while its declarative programming paradigm enables developers to build complex user interfaces more efficiently and predictably. The component-based architecture of React promotes code modularity and reuse, providing development flexibility(See details in Fig. 5).

For backend services, the platform adopts the Spring Boot framework, an open-source Java-based framework designed for creating independent, production-level Spring applications. Spring Boot's automatic configuration and modular design simplify the startup and operation processes of Spring applications. It also features an embedded web server and provides various production-ready features, including health checks and external configuration support, ensuring stability and maintainability. In the backend, the large model inference service uses the LLMs framework, which manages key-value memory in the attention mechanism via Paged Attention. This enables rapid execution of models using CUDA/HIP graph and supports tensor and pipeline parallelism for distributed inference, boosting the inference speed of large models.

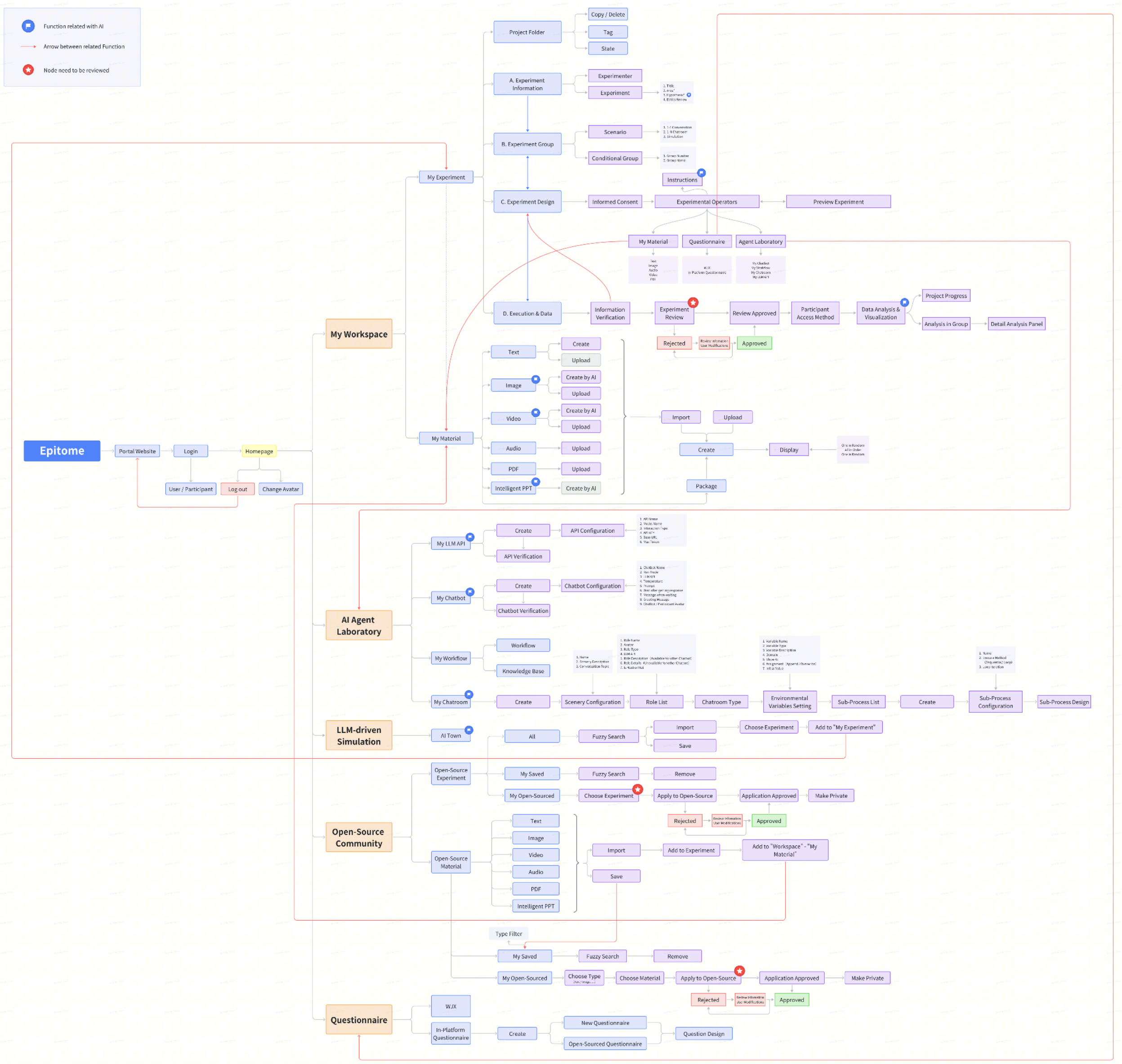

**Fig.5|Product Structure and User Flow of Epitome web platform.**

**Four-Tier Framework for Epitome's Ethical Safeguard Strategy.** *Epitome* incorporates ethical safeguards aligned with principles of responsible AI research and human subjects protection. The platform implements a four-tier framework: (1) informed consent mechanisms that clearly disclose when participants interact with AI agents; (2) data minimization protocols that collect only information essential to research questions; (3) transparency features that allow participants to review and understand AI-generated responses; and (4) oversight mechanisms that facilitate IRB review of experimental designs. These safeguards address the unique ethical challenges of human-AI experiments, particularly regarding participant autonomy and potential manipulation concerns (see details in Fig. 6).

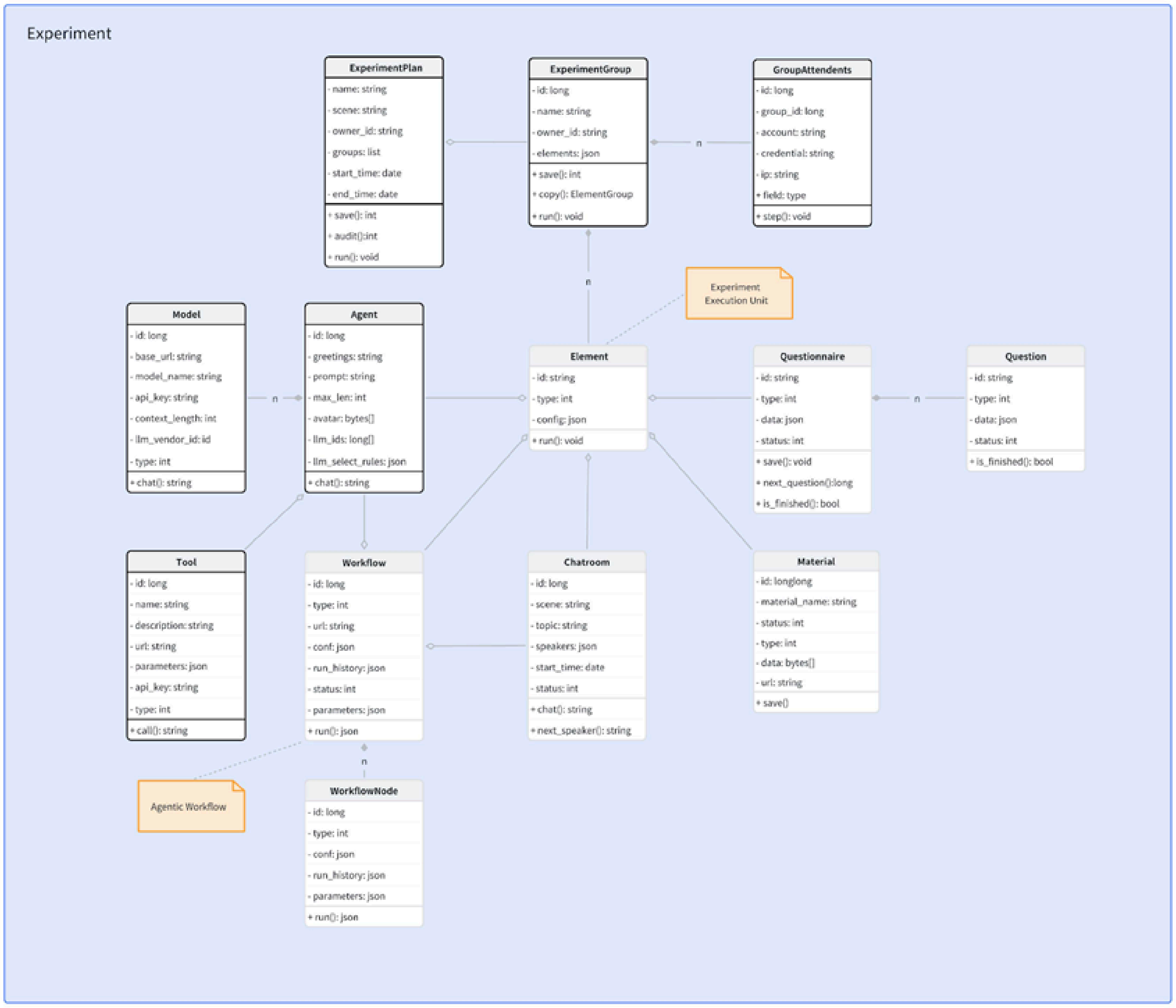

**Fig.6|Unified Modeling Language figure for Epitome layout.**

## Data, Materials, Software Availability

The detailed data of the case studies in the paper and related materials associated with the experiments are available in Supplementary Information. For more and latest information, refer to [GitHub - epitome-AISS/epitome](GitHub - epitome-AISS/epitome). Moreover, a user manual for all experiment designers is provided, offering detailed guidance on the functions and operational procedures of the *Epitome* platform (see Supplementary Section G). *Epitome* platform now uses a managed access model, interested researchers who wish to use *Epitome* to support their experiments are encouraged to contact the authors for access. For application or accession, refer to www.epitome-ai.com.